\documentclass[prd,amsmath,amssymb,preprint]{revtex4}
\usepackage{graphicx}
\usepackage[english]{babel}
\usepackage{dcolumn}
\usepackage{bm}
\usepackage{longtable,booktabs}
\usepackage{rotating}
\usepackage{afterpage}
\usepackage{pdflscape}
\usepackage{multirow}
\usepackage{hyperref}
\usepackage{siunitx}
\usepackage{booktabs}
\usepackage{amsmath}
\usepackage{physics}

\usepackage{color}
\definecolor{BLUE}{rgb}{0.0,0.0,1.0}
\definecolor{BLACK}{rgb}{0.0,0.0,0.0}

\usepackage{soul}
\soulregister\cite7
\soulregister\ref7
\newcommand{\rs}[1]{}

\begin{document}
\title{Relativistic calculations of the ground and inner-L-shell excited energy levels of berylliumlike ions}
\author{M.~Y.~Kaygorodov$^{1}$, Y.~S.~Kozhedub$^{1,2}$, 
I.~I.~Tupitsyn$^{1,3}$, A.~V.~Malyshev$^{1,2}$, D.~A.~Glazov$^{1}$, 
G.~Plunien$^{4}$, and V.~M.~Shabaev$^{1}$
}
\affiliation{
$^1$Department of Physics, St. Petersburg State University,
7/9 Universitetskaya nab., 199034 St.~Petersburg, Russia\\
$^2$ NRC ``Kurchatov Institute'', Academician Kurchatov 1, 123182 Moscow, 
Russia\\
$^3$ Center for Advanced Studies, Peter the Great St. Petersburg Polytechnic 
University, Polytechnicheskaya 29, 195251 St. Petersburg, Russia\\
$^4$Institut f\"ur Theoretische Physik, TU Dresden, Mommsenstrasse 13, 
D-01062 Dresden, Germany}
\begin{abstract}
Large-scale relativistic configuration-interaction method combined with
many-body perturbation theory is consistently applied to calculations of the energy 
levels of the ground and  inner-L-shell excited states
of berylliumlike ions in the range $10 \leq Z \leq 92$. 
The quantum electrodynamics, nuclear recoil, and frequency-dependent Breit 
corrections are taken into account. 
The obtained results are supplemented with the systematical estimation of the uncertainties.
\end{abstract}
          

\maketitle
\section{Introduction}

The development of high-precision physics of few-electron heavy ions 
in recent decades was motivated by testing quantum electrodynamics (QED) at strong fields, precise determination of the fundamental constants and searches for their time variation, provision of astrophysical investigations and other applications, see, e.g., the recent reviews~\cite{RevModPhys.90.025008,RevModPhys.90.045005,shabaev:18:hi} and references therein. The most advanced calculations of the binding energies of highly-charged ions are performed within \textit{ab initio} QED approach and include all the corrections up to the second-order QED contributions.  
To date, these state-of-the-art calculations have been done for the ground and low-lying excited states of He-like~\cite{Lindgren1996PRL,Yerokhin1997PRA_He-like_2eSE, Artemyev2005PRA_He-like,Malyshev:2019:PRA:accepted}, Li-like~\cite{YerokhinPRA2001_Li-like,Kozhedub2010PRA,Sapirstein2011PRA_Li-like}, and B-like~\cite{Artemyev2007PRL_B-like,Artemyev2013PRA_B-like,Malyshev:2017:022512} ions. For Be-like ions, however, the QED calculations of this kind have been accomplished for the ground state only~\cite{Malyshev2014PRA_Be-like_GS,Malyshev2015PRA_Be-like}. The energies of the excited states of Be-like ions were studied previously within the various relativistic methods~\cite{Safronova1996PRABe-like, Chen1997PRA_Be-like, Safronova2000MolPhys_Be-like, Majumder2000PRA_Be-like_M2, GuADNDT2005, Ho2006PRA_Be-like, Cheng2008PRABe-like, Jian-Hui2008CommThPhys_Be-like, Sampaio2013PhysScr_Be-like_Kr, Yerokhin2014PRA_Be-likeFe, Yerokhin2015PhysScr_Be-likeAr}.

The main goal of the present work is a high-precision evaluation of the electron-electron correlation effects for the ground and inner-L-shell excited states of Be-like ions within the Breit approximation. 
Berylliumlike ions are the simplest examples of atomic systems where both intrashell and intershell interactions are important.
High-precision theoretical predictions require accurate evaluation of both kinds of interaction.
To perform the calculations we employ the large-scale relativistic configuration-interaction (CI) method combined with many-body perturbation theory (PT) in the basis of Dirac-Sturm orbitals.
Special attention is payed to a systematic estimation of the uncertainties of the obtained results. 
In order to obtain the total energy-level values, the CI energies are supplemented by the frequency-dependent Breit and nuclear recoil corrections. The radiative QED effects are taken into account by means of the model QED operator approach~\cite{Shabaev2013PRA_ModelOperator,Shabaev2015QEDMOD}. 
The calculations are carried out along the beryllium isoelectronic sequence with the nuclear charge number in the range $10 \leq Z \leq 92$.
All these data yield predictive results, i.e., the results with systematical estimations of numerical errors and uncertainties due to uncalculated effects.
The obtained theoretical predictions of the transition energies provide a straightforward way for improving as soon as the higher-order QED corrections will be available.

The paper is organized as follows. In the next section, we give a brief outline of our approach and computation method.
Section~\ref{sec:discussion} presents the numerical results and the comparison with the previous theoretical calculations and available experimental data.
The relativistic units $(\hbar = c =1)$ and the Heaviside charge unit $(\alpha = e^2/4\pi,\, e < 0)$ are used throughout the paper.

\section{Methods of calculations}
%
\subsection{Electron correlations}
{\color{black} Within the lowest-order relativistic approximation, the energy $E$ of an $N$-electron atom is given by the eigenvalue of the Dirac-Coulomb-Breit (DCB) equation
\begin{equation}\label{DCB_equation}
H_{\mathrm{DCB}} \Psi = E \Psi,
\end{equation}
where $\Psi$~is the many-electron wave function.}
The DCB Hamiltonian in the ``no-pair'' approximation is 
conventionally defined by 
%
\begin{gather}
\label{DCB_Hamiltonian}
H_{\mathrm{DCB}} = \Lambda^{(+)} \left[ H_0 + V_{\rm int} \right] \Lambda^{(+)},  \\
\label{DCB_Hamiltonian_H0_Vint}
H_0 = \sum_i^N h_{\rm D}{(i)}, \qquad 
V_{\rm int} = \sum_{i<j}^N \left[ V_{\mathrm{C}}{(i,j)} + 
V_{\mathrm{B}}{(i,j)}\right] \, ,
\end{gather}
where the indices $i$ and $j$ numerate the electrons, $h_{\rm D}$ is the one-particle Dirac Hamiltonian, 
\begin{equation}\label{h_one_el}
h_{\rm D} = \pmb{\alpha} \cdotp \pmb{p} 
 + (\beta - 1)m + V_{\rm nucl}(r),
\end{equation}
$\pmb{\alpha}$ and $\beta$ are the Dirac matrices, 
$\pmb{p}$ is the momentum operator, 
$m$ is the electron mass,
$V_{\rm nucl}$ is the binding potential of the nucleus, $V_{\mathrm{C}}$ and $V_{\mathrm{B}}$ are the 
Coulomb part and the frequency-independent Breit part of the electron-electron 
interaction, respectively,
\begin{equation}\label{V_coul}
V_{\mathrm{C}}{(i,j)} = \frac{\alpha}{r_{ij}},
\end{equation}
\begin{equation}\label{V_br}
V_{\mathrm{B}}{(i,j)} = -\frac{\alpha}{2r_{ij}}
\left[ \pmb{\alpha}_i \cdot \pmb{\alpha}_j 
+ \frac{(\pmb{\alpha}_i \cdot \pmb{r}_{ij})(\pmb{\alpha}_j \cdot 
\pmb{r}_{ij})}{r_{ij}^2} \right].
\end{equation}
Here $\alpha$ is the fine structure constant, $r_{ij} = |\pmb{r}_{ij}| = |\pmb{r}_{i}-\pmb{r}_{j}|$, $\pmb{r}_{i}$ is the position vector of the $i$-th electron. 
$\Lambda^{(+)}$ in Eq.~(\ref{DCB_Hamiltonian}) is the product of the 
one-electron projectors on the positive-energy states which correspond to 
some one-particle Dirac Hamiltonian $\tilde{h}_{\rm D}$, which does not 
necessarily coincide with~(\ref{h_one_el}).
{\color{black} In principle, the Hamiltonian~$\tilde{h}_{\rm D}$ may include even nonlocal parts, e.g., the Dirac-Fock-Sturm operator.  However, having in mind that the results of the present work are supposed to be merged with the anticipated rigorous QED calculations, in the present work we define the operator~$\tilde{h}_{\rm D}$ as follows
\begin{equation}\label{h_one_el_scr}
\tilde{h}_{\rm D} = \pmb{\alpha} \cdotp \pmb{p} 
 + (\beta - 1)m + V_{\rm nucl}(r) + V_{\rm scr}(r),
\end{equation}
where $V_{\rm scr}(r)$ is the local screening potential which models the interelectronic interaction. Varying the screening potential~$V_{\rm scr}(r)$ in Eq.~(\ref{h_one_el_scr}) modifies the definition of the positive-energy-states projection operator $\Lambda^{(+)}$ and, therefore, slightly influences the solutions of the DCB equation~(\ref{DCB_equation}). This small ambiguity in the DCB results is determined by the arbitrariness of the realization of the ``no-pair'' approximation and can be fully eliminated only within the rigorous bound-state QED approach. The uncertainty related to the ambiguity is kept under control in the present work.
}

In the present study, we treat the correlations within the framework of the large-scale CI approach with the configuration-state-function space which includes the quadruple excitations from the reference state(s).
In order to obtain the high-precision DCB energies, we combine this approach with the many-body PT and the CI+PT~\cite{Dzuba1996PRA_CI+MPBT,Kozlov:2015:199} methods.
High-precision calculations must take into account all types of the correlations: the valence-valence, the core-valence, and the core-core ones.
Within the CI+PT method, the configuration-interaction calculations are carried out for the valence electrons while the residual core-valence and core-core interaction is treated by means of the many-body perturbation theory taking into account the contributions up to the third order.
In contrast to the CI approach, for a given set of the configuration-state functions the PT and CI+PT methods provide only an approximate solution of Eq.~(\ref{DCB_equation}), treating the interelectronic-interaction effects incompletely. 
However, PT and CI+PT calculations can be performed employing the much more saturated basis sets and, therefore, provide corrections, which can be used to improve the final CI results.
{\color{black}The PT series in the present work are constructed employing the initial approximation, which is chosen to be consistent with the definition of the Hamiltonian~(\ref{h_one_el_scr}). That is, the zeroth-order approximation in PT calculations corresponds to the Hamiltonian 
$\sum_i^N \tilde{h}_{\rm D}{(i)}$.}
The numerical scheme is discussed in more details in Sec.~\ref{sec:discussion}.

Another important feature of the present evaluations is a systematic estimation of the numerical uncertainties of the obtained results.
For each particular state and each nuclear charge, we perform the calculations with a large number of the different sets of the configuration-state functions.
By analyzing the successive increments of the results obtained with the increasing basis sets, we conclude how well the PT, CI+PT and the final CI results converge. 

\subsection{QED effects} \label{ssc:qed_effects}

Accurate theoretical predictions of the energy levels demand a supplementation of the relativistic energies obtained from the DCB Hamiltonian by the QED corrections.
Because of significant technical difficulties, \textit{ab initio} calculations of the many-electron QED effects, in particular, for berylliumlike ions, are mainly restricted to the ground state. 
For excited states, especially for quasi-degenerate states, or more complex atoms (ions) one has to rely on a simplifed treatment of the QED effects.


First, we should take into account the frequency-dependent Breit correction.
This correction can be obtained by using the full Coulomb-gauge QED interelectronic-interaction operator (see, e.g. Ref. \cite{Shabaev2013PRA_ModelOperator} and references therein) instead of the Coulomb $V_\mathrm{C}$ and standard Breit $V_\mathrm{B}$ interactions in Eq.~(\ref{DCB_Hamiltonian}).
Following Ref.~\cite{Yerokhin2012PRA_Li-like}, in order to avoid the spurious effects, we perform this substitution only for the DCB Hamiltonian matrix elements with the reference-state wave functions.
The uncertainty due to the omitted higher-order QED electron-correlation contributions is conservatively estimated to be $\pm\alpha^2 (\alpha Z)^3 m$~r.u. or $\pm (\alpha Z)^3$~a.u..
According to the available data for helium-, lithium-, and berylliumlike ions this estimation appears to be reliable.

The self-energy and vacuum-polarization corrections are treated by means of the model QED (QEDMOD) operator $h_{\mathrm{QED}}$~\cite{Shabaev2013PRA_ModelOperator,Shabaev2015QEDMOD} included into the DCB Hamiltonian (\ref{DCB_Hamiltonian}).
Taking the difference between the CI energies evaluated with and without $h_{\mathrm{QED}}$, we obtain the QED correction to the total energy. 
If the evaluation is carried out with the Dirac-Coulomb basis set, this approach takes into account the QED effects to zeroth order in $1/Z$ completely. 
The screened QED contributions of the first and higher orders in $1/Z$ are treated in this way only approximately. 
In order to estimate the accuracy of the method we compared the screening effects evaluated employing the QEDMOD approach with the rigorous QED calculations of the first order in the interelectronic interaction for He-like ions~\cite{Artemyev2005PRA_He-like}, Li-like ions~\cite{Kozhedub2010PRA} and the available states of Be-like ions~\cite{Malyshev2014PRA_Be-like_GS,Malyshev2015PRA_Be-like,Yerokhin2014PRA_Be-likeFe,Yerokhin2015PhysScr_Be-likeAr}.
Finally, we estimate the uncertainty of the screened QED effects calculated to be at the level of $15\%$ for neon and growing linearly up to $30\%$ for uranium.
This also includes the uncertainty due to omitted two-loop one-electron QED effects.

\subsection{Nuclear recoil effect}
The nuclear recoil correction to the DCB energy accounts for the finite mass of the nucleus.
The fully relativistic theory of the recoil effect can be formulated only within QED~\cite{Shabaev1985TMP_Recoil,Shabaev1988,Shabaev1998PRA_QEDRecoil,Adkins:2007:042508}.
In the lowest-order relativistic approximation and in the first order in the electron-to-nucleus mass ratio $m/M$ the nuclear recoil effect on the binding energy can be treated exploiting the following mass shift (MS) Hamiltonian~\cite{Shabaev1985TMP_Recoil,Shabaev1988, Palmer1987}
\begin{equation}\label{MS_hamiltonian}
H_{\mathrm{MS}} = \frac{1}{2M}\sum_{i,j} \left\lbrace \bm{p}_i \cdot \bm{p}_j - 
\frac{\alpha Z}{r_i} \left[ \bm{\alpha_i} + \frac{( \bm{\alpha}_i \cdot 
\bm{r}_i)\bm{r}_i}{r^2_i} \right] \cdot \bm{p}_j \right\rbrace.
\end{equation}
In the present paper, the energy shifts due to the nuclear recoil effect are evaluated as the expectation value of the MS Hamiltonian (\ref{MS_hamiltonian}) with the solutions of the DCB equation (\ref{DCB_equation})~\cite{Tupitsyn2003PRA_IS},
\begin{equation}\label{E_MS_contrubution}
\Delta E_{\mathrm{MS}} = \expval{H_{\mathrm{MS}}}{\Psi}.
\end{equation}

The QED nuclear recoil corrections can be comparable in magnitude to the results obtained within the lowest-order relativistic approximation for high-$Z$ ions, see, e.g., Ref.~\cite{Artemyev:1995:1884,Artemyev:1995:5201,Shabaev:1999:493,Malyshev:2018:085001}. 
In the present work, we omit these corrections since they are negligible compared to the uncertainty of the other QED contributions. 
The QED nuclear recoil effect on the binding energies of Be-like ions, in principle, can be estimated within the independent-electron approximation using the results presented in Ref.~\cite{Malyshev:2018:085001}.

\section{Results and discussion}\label{sec:discussion}

In the present work, the numerical procedure of solving the DCB equation 
(\ref{DCB_equation}) is based on the CI method in the basis of the four-component Dirac-Sturm (DS) orbitals~\cite{Tupitsyn2003PRA_IS, Tupitsyn2005PRA_Be-likeB-like_M1}. 
The employed one-electron basis set consists of the eigenfunctions of the Dirac equation {\color{black}with the Hamiltonian~(\ref{h_one_el_scr})} for the K and L shells and the solutions of the {\color{black} corresponding} Dirac-Sturm equation for the other orbitals. 
Then, the projection operator $\Lambda^{(+)}$ in Eq.~(\ref{DCB_Hamiltonian}) is constructed from the positive-energy eigenvectors obtained by the diagonalization of the one-electron Dirac Hamiltonian $\tilde{h}_{\rm D}$ in this basis. 
All the calculations are performed with the local Dirac-Fock (LDF)~\cite{Shabaev2005PRA_NPC_Fr&Cs} screening potential included into the Dirac and Dirac-Sturm equations.
Introducing a screening potential in the zeroth-order approximation implies that the PT series are constructed for the residual interelectronic interaction.
The LDF potential is designed by the inversion of the radial Dirac equation with the radial wave functions obtained in the Dirac-Fock approximation.
To describe the nuclear charge distribution, the standard Fermi model with the nuclear radii taken from Ref.~\cite{Algeli2013ADNDT} is used.
We remind that the key condition for merging the CI results with \textit{ab initio} QED calculations is the use of the consistent projection operator $\Lambda^{(+)}$, while the employed one-electron basis set could be obtained for any other screening potential or, e.g., nonlocal Dirac-Fock-Sturm operator.

The direct application of the CI method for Be-like ions leads to time-consuming computational task which demands significant resources to be used.
Meanwhile, the major part of the interelectronic interaction can be captured by means of the lowest orders of the many-body PT which allows one to perform the calculations with the larger basis sets at the same cost.
Therefore, in order to reach high accuracy, in the present work the CI method was combined with the many-body PT (up to the third order) and the CI+PT approaches~\cite{Dzuba1996PRA_CI+MPBT}. 
The numerical scheme to evaluate the DCB energies can be briefly described as follows. 
At the first stage, we perform the calculations of the second order of the PT taking into account all kinds of the single (S) and double (D) excitations (PT2~SD) (the zeroth (PT0) and first (PT1) orders are trivial).
In Table~\ref{table:table1}, the contributions of the second order of the PT to the ground-state energy of berylliumlike neon ($Z=10$) are presented as an example.
We carry out a series of calculations with different bases and analyze the successive increments as the basis is enlarged.
In our analysis, we study the dependence of the obtained results on the parameter $L_{\mathrm{max}}$, which is the maximum value of the orbital quantum number for the DS orbitals, and the parameter $N_{\rm max}$, which is the number of the DS orbitals for each $L$.
By extending the basis set and taking the differences between the results, we identify the contributions of the individual partial waves and check the stability of the results for each orbital number $L$ with regard to the number of the basis functions.
The $L_{\rm max}\to \infty$ extrapolation is done by polynomial least-square fitting in $1/L$. 
Finally, the extrapolation to the case $N_{\rm max}\to \infty$ is performed in the same way.

In Table~\ref{table:table2}, in order to illustrate the employed numerical scheme we present the detailed results for the calculations of the DCB energies for the ground and inner-L-shell excited states in Ne$^{6+}$, Xe$^{50+}$, and U$^{88+}$ ions.
The PT0 values are given in the first line for each ion. All the other rows represent various corrections to the PT0 values. These corrections are obtained by the successive application of the more and more sophisticated approaches to treat the correlation effects. 
The notations for the corresponding methods are indicated in the second column of Table~\ref{table:table2}. 
In the following, we use the same notations for the methods and the corrections obtained employing these methods.

For instance, the term labeled with ``PT3~SD'' in Table II is obtained by performing the PT calculations up to the third order including the SD excitations and subtracting the PT0, PT1 and PT2~SD values evaluated using the same basis. 
Since the Rayleigh-Schr\"odinger PT is employed, the PT3~SD term is just the PT contribution of the third order.
Having performed the PT calculations, at the next stage we apply the CI and CI+PT methods in order to take into account the higher-order interelectronic-interaction effects. 
First, we restrict the configuration space to the SD excitations only.
In the CI+PT method the configuration space is divided into a smaller subspace for the CI calculations and its orthogonal complement which is treated perturbatively.
When the smaller CI subspace is constructed, only the excitations from the $n=2$ orbitals of the reference state(s) into the virtual shells with $2\leqslant n\leqslant 6$ are considered. 
Having completed the CI calculations, the interaction with the orthogonal complement is evaluated by the PT taking into account all the contributions up to the third order.
To extract the correction to the contributions calculated at the previous stage, we take the difference between the results obtained for the identical basis within the CI+PT3~SD and PT3~SD approaches. 
The corresponding corrections are labeled ``CI+PT3~SD'' in Table~\ref{table:table2}. 
The CI+PT3~SD term provides the correction to the PT energy which is induced by the most important configurations. 
After this, employing the identical configuration spaces and considering the difference of the values calculated within the CI and CI+PT3 approaches, we extract the CI contribution for the configurations which hold the highly-excited virtual orbitals. 
The latter correction is indicated as ``CI~SD'' in Table~\ref{table:table2}. 
Then we gradually expand the configuration space of the initial problem by considering the triple (T) and quadruple (Q) excitations.
Following the same procedure, we obtain the CI+PT3~T, CI~T, CI+PT3~Q, and CI~Q corrections, see Table~\ref{table:table2}. 
Finally, summing up all the contributions listed in Table~\ref{table:table2} we obtain the total value of the DCB energy for a given state.

In our CI+PT calculations, namely the~CI+PT3~SD, CI+PT3~T, and CI+PT3~Q, we limit the basis set with ($L_{\mathrm{max}} = 5$, $N_{\rm max}=25$).
Since the nonperturbative treatment of the $1s$ orbitals and the highly-excited virtual orbitals considerably increases the configuration space\, we restrict the basis sets parameters for CI~T and CI~Q calculations to ($L_{\mathrm{max}} = 3$, $N_{\rm max}=25$) and ($L_{\mathrm{max}} = 2$, $N_{\rm max}=25$), respectively.
From Table~\ref{table:table2} one can see that the triple and quadruple 
excitations are important only for the ground state and are almost negligible for the inner-L-shell excited states.
The uncertainties of all the contributions are mainly determined by the estimation of the convergence of the results with respect to the maximum value of the orbital quantum number $L_{\mathrm{max}}$ for the DS orbitals. 

In order to illustrate the dependence of the obtained DCB energies on the realization of the ``no-pair'' approximation (see the discussion after Eq.~(\ref{DCB_Hamiltonian})) we have performed the calculations for Be-like neon and uranium starting from the different initial approximations. 
In addition to the LDF potential introduced in Eq.~(\ref{h_one_el_scr}), we have evaluated the DCB energies for the core-Hartree potential induced by the $1s^2$ closed shell as well.
The results for the DCB energies without and with the frequency-dependent Breit correction are presented in Table~\ref{table:table3}.
One can see that for high-$Z$ ions the scatter of the results obtained for the different projection operators $\Lambda^{(+)}$ in Eq.~(\ref{DCB_Hamiltonian}) is higher that the numerical uncertainty.
For low-$Z$ ions this is not the case and the results obtained for the different initial approximations lie within the estimated error bar.
From Table~\ref{table:table3}, it is seen that the frequency-dependent Breit correction can not eliminate this discrepancy between the results.
This issue can be solved only by considering the excitations into the negative-energy continuum within the rigorous QED approach.
In the following, this ambiguity of the DCB results is kept under control and covered by our estimation of the higher-order QED electron-correlation contributions which is shown in the last column of Table~\ref{table:table3}.


In Table~\ref{table:table4} we present the results for the DCB energies of the ground and inner-L-shell excited states of Be-like ions with the frequency-dependent Breit correction included (the DCB energies without this correction can be found in third column of Table~\ref{table:table5}). 
For the ground state the total energies are listed whereas for the excited states the energies relative to the ground state are given.
The presented uncertainties are of purely numerical origin. 
Sometimes the behavior of the uncertainties is not smooth enough when nuclear charge $Z$ changes. 
As it was noted above, the error bar is determined mainly by the 
convergence of the PT, which is a little bit better for some $Z$ and states and a little bit worse for the others. 
We intentionally do not use any correction procedure and suppose that the uncertainty is reliable anyway.      
There exist many nonrelativistic and relativistic calculations of the correlation effects for berylliumlike ions in the literature~\cite{Safronova1996PRABe-like,Chen1997PRA_Be-like,Safronova2000MolPhys_Be-like,Majumder2000PRA_Be-like_M2,GuADNDT2005,Ho2006PRA_Be-like,Cheng2008PRABe-like,Jian-Hui2008CommThPhys_Be-like,Sampaio2013PhysScr_Be-like_Kr,Li2017CanJPhys_Be-like}. 
We compare our results with the most accurate ones: the CI calculations by Yerokhin \textit{et al.} for Be-like argon~\cite{Yerokhin2015PhysScr_Be-likeAr} and iron~\cite{Yerokhin2014PRA_Be-likeFe}, the CI calculations by Cheng \textit{et al.}~\cite{Cheng2008PRABe-like}, the many-body PT calculations by Safronova \textit{et al.}~\cite{Safronova1996PRABe-like,Safronova2000MolPhys_Be-like}, and our previous \textit{ab initio} QED evaluation of the ground-state energy~\cite{Malyshev2014PRA_Be-like_GS}.
We find our results to be in reasonable agreement with the previous ones, and have better numerical accuracy.
Since the authors use different ``no-pair'' approximations, one can not expect to obtain the agreement better than the corresponding ambiguity.

Finally, in Table~\ref{table:table5} we collect the DCB, frequency-dependent Breit, QEDMOD, and nuclear recoil contributions to the total energies of the ground and inner-L-shell excited states of Be-like ions. 
As in Table~\ref{table:table4}, for the ground state the total energies are given while for the excited states the energies relative to the ground state are shown.
The DCB energies are presented with the numerical uncertainty (in the first brackets) as well as with the uncertainty due to the finite nuclear size effect (in the second brackets). 
The last one is conservatively estimated by adding quadratically two contributions. 
The first contribution is obtained by varying the root-mean-square nuclear radius within the error bars given in Ref.~\cite{Algeli2013ADNDT}, and the second one is obtained by changing the model of the nuclear-charge distribution from the Fermi one to the homogeneously-charged-sphere model.
In the case of uranium, we also take into account the nuclear deformation effect in accordance with the results of Ref.~\cite{Kozhedub2008PRA}.
The column labeled ``Breit($\omega)$'' represents the frequency-dependent Breit contribution.
The next two columns contain the QEDMOD and nuclear recoil contributions, respectively.
The final uncertainty is mainly determined by the omitted higher-order QED electron-correlation contributions, see the discussion in subsection~\ref{ssc:qed_effects}. Its estimation is included in the uncertainty of the ``Total'' values in addition to the previously discussed ones.
The obtained results are in reasonable agreement with the previous calculations, the data from the NIST Atomic Spectra Database~\cite{NIST_ASD}, and the available experimental values \cite{Edlen1983PhysScr_Be-like,Bernhardt2015JPhysB_Be-like_Xe}. We consider our calculations as the best ones available, and our conservative estimation of the uncertainty can only be reduced by rigorous evaluation of the higher-order QED contributions.


\section{Conclusion}

To summarize, we have performed the systematic relativistic calculations of the ground and  inner-L-shell excited energy levels of berylliumlike ions from neon to uranium. 
The main feature of the work is the high-precision evaluation of the Dirac-Coulomb-Breit contribution based on the large-scale relativistic configuration-interaction method combined with the perturbation theory and are supplemented with the systematic estimation of the uncertainties. 
In order to estimate the uncertainties of the CI energies, we have thoroughly analyzed the successive increments of the results obtained with the sets of configuration-state functions increased in all possible directions. 
The DCB energies are combined with the separately computed corrections: frequency-dependent Breit, QED (using the model QED operator), and nuclear recoil (using the relativistic mass shift operator).
All our theoretical predictions are given with the uncertainties that include the estimation of the effects omitted in the present theoretical treatment.
For the energy of the ground state, our results agree well with the benchmark theoretical data available in the literature \cite{Malyshev2014PRA_Be-like_GS}.
For the transition energies, our theoretical approach provides the most precise values for berylliumlike ions in the range $Z=10$--$92$.
For Be-like argon and iron our estimation of the omitted QED contributions is performed in a more conservative way than in Refs.~\cite{Yerokhin2014PRA_Be-likeFe, Yerokhin2015PhysScr_Be-likeAr}.
The total accuracy for all $Z$ is determined by the uncalculated contributions of the higher-order QED effects and can be improved as soon as these corrections are calculated. 

\section{Acknowledgment}

We thank Vladimir Yerokhin for valuable discussions. 
This work was supported by RFBR (Grants No.~16-02-00334, No.~18-03-01220), SPbSU-DFG (Grants No.~11.65.41.2017 and No.~STO~346/5-1) and by
Ministry of Education and Science of the Russian Federation
(Grant No. 3.1463.2017/4.6). 
Y.S.K. and A.V.M. acknowledge the financial support from the FAIR-Russia Research Center. 
The work of Y.S.K. and D.A.G. was also supported by SPbSU (COLLAB 2018 id 34827923 and id 34824940). 
M.Y.K., A.V.M., and V.M.S. acknowledge the financial support from the Foundation for the advancement of theoretical physics and mathematics ``BASIS''.
The research was carried out using the computational resources provided by the St.~Petersburg State University Computing Center.

\pagebreak

\begin{table}
\caption{The contributions of the second order of the perturbation theory to the Dirac\,-\,Coulomb\,-\,Breit ground-state binding energy of Ne$^{6+}$, in a.u.. 
The calculations are performed for the LDF potential including the single and double excitations. 
The values listed after the first row are the increments obtained by adding successively the configurations while the maximal value of the orbital quantum number $L_{\mathrm{max}}$ is increased. 
The columns from left to right correspond to the enlargement of the size of the one-electron basis set, and $N_{\mathrm{max}}$ labels the number of the one-electron virtual orbitals for a particular $L_{\mathrm{max}}$.
The last line shows the total results of the calculations with the particular basis sets.
The final value is obtained by the extrapolation to the case $N_{\rm max}\to\infty$.}
\label{table:table1}
\centering
\resizebox{\columnwidth}{!}
{
\begin{tabular}{c@{\qquad}
                  S[table-format=-2.6(2)] 
                  S[table-format=-2.6(2)] 
                  S[table-format=-2.6(2)] 
                  S[table-format=-2.6(2)] 
                  S[table-format=-2.6(2)] 
                  S[table-format=-2.6(2)]}
\hline
\hline
$L_{\mathrm{max}}$
&\multicolumn{1}{c}{$N_{\rm max}=10$}
&\multicolumn{1}{c}{$N_{\rm max}=15$}
&\multicolumn{1}{c}{$N_{\rm max}=20$}
&\multicolumn{1}{c}{$N_{\rm max}=25$}
&\multicolumn{1}{c}{$N_{\rm max}=30$}
&$N_{\rm max}\to \infty$\\
\hline
1	&-0.232241	&-0.232472	&-0.232481	&-0.232483	&-0.232484\\
2	&-0.011581	&-0.011632	&-0.011636	&-0.011638	&-0.011638\\
3	&-0.003028	&-0.003063	&-0.003069	&-0.003070	&-0.003071\\
4	&-0.001111	&-0.001144	&-0.001149	&-0.001151	&-0.001151\\
5	&-0.000495	&-0.000521	&-0.000526	&-0.000526	&-0.000528\\
6	&-0.000247	&-0.000268	&-0.000273	&-0.000276	&-0.000275\\
7	&-0.000138	&-0.000152	&-0.000157	&-0.000158	&-0.000159\\
8	&-0.000082	&-0.000091	&-0.000095	&-0.000097	&-0.000098\\
9	&-0.000052	&-0.000059	&-0.000062	&-0.000063	&-0.000064\\
10	&-0.000034	&-0.000039	&-0.000042	&-0.000043	&-0.000043\\
$11-\infty$	&-0.000117(24)	&-0.000133(21)	&-0.000144(22)	&-0.000151(22)	&-0.000156(23)\\
Total	&-0.249126(24)	&-0.249574(21)	&-0.249634(22)	&-0.249656(22)	&-0.249667(23) &-0.249685(29)\\
\hline
\hline
\end{tabular}
}
\end{table}

\begin{table}
\caption{Contributions to the DCB energy of the ground and singly-excited $n=2$ states in Ne$^{6+}$, Xe$^{50+}$, and U$^{88+}$, in a.u.. The calculations are performed for the LDF potential. See the text for details.}
\label{table:table2}
\centering
\resizebox{\columnwidth}{!}
{
\begin{tabular}{l l S[table-format=-5.6(3)]
                    S[table-format=-5.6(2)]
                    S[table-format=-5.6(2)] 
                    S[table-format=-5.6(3)] 
                    S[table-format=-5.6(2)]}
\hline
\hline
Ion
&Contribution
&\multicolumn{1}{c}{$1s^22s^2$ $^1S_0$}
&\multicolumn{1}{c}{$1s^22s2p$ $^3P_0$}
&\multicolumn{1}{c}{$1s^22s2p$ $^3P_1$}
&\multicolumn{1}{c}{$1s^22s2p$ $^1P_1$}
&\multicolumn{1}{c}{$1s^22s2p$ $^3P_2$}\\
\hline
Ne$^{6+}$
&PT0		&-91.402481	        &-90.768504         &-90.768504	        &-90.760454	        &-90.760454\\
&PT1		&-18.825843	        &-19.065206         &-19.063164	        &-18.547442	        &-19.066726\\
&PT2 SD		&-0.249685(29)	    &-0.085084(13)      &-0.085063(13)  	&-0.143987(71)	    &-0.085042(15)\\
&PT3 SD		&0.066395(74)		&0.003883(9)    	&0.003877(8)	    &0.009919(59)		&0.003883(8)\\
&CI+PT3 SD	&-0.006567(46)		&0.000008(4)	    &0.000008(3)	    &0.000064(49)		&0.000008(3)\\
&CI SD		&-0.000399(30)		&-0.000193(4)	    &-0.000190(4)	    &-0.000244(44)		&-0.000190(5)\\
&CI+PT3 T	&-0.001555(8)		&-0.000042(3)	    &-0.000042(3)	    &-0.000349(16)		&-0.000042(3)\\
&CI T		&0.000063(9)		&-0.000019(3)	    &-0.000019(2)	    &-0.000024(7)		&-0.000018(2)\\
&CI+PT3 Q	&-0.002688(22)		&-0.000002(1)	    &-0.000002(1)	    &-0.000010(5)		&0.000002(1)\\
&CI Q		&0.000069(19)		&0.000000		    &0.000000		    &0.000000			&0.000000\\
&Total		&-110.422691(102)	&-109.915159(17)	&-109.913099(16)    &-109.442527(115)	&-109.908582(18)\\
\hline
Xe$^{50+}$
&PT0		&-3597.287490   	&-3592.603624	    &-3592.603624	    &-3578.750225	    &-3578.750225\\
&PT1		&-116.550483    	&-117.521971	    &-116.671166	    &-115.601314	    &-117.977435\\
&PT2 SD		&-0.574620(38)  	&-0.136988(42)	    &-0.150397(38)	    &-0.179117(55)	    &-0.131641(39)\\
&PT3 SD		&0.098523(39)		&0.001185(13)		&0.001507(18)		&0.002115(23)		&0.001099(12)\\
&CI+PT3 SD	&-0.008335(8)		&0.000001(1)		&-0.000041(5)		&0.000043(10)		&0.000001(1)\\
&CI SD		&-0.000041(5)		&-0.000009(2)		&-0.000012(4)		&-0.000009(3)		&-0.000009(4)\\
&CI+PT3 T	&-0.000836(4)		&-0.000002(1)		&-0.000005(2)		&-0.000013(3)		&-0.000002(1)\\
&CI T		&0.000011(4)		&-0.000002(1)		&-0.000002(2)		&-0.000002(2)		&-0.000002(1)\\
&CI+PT3 Q	&-0.001627(5)		&0.000000			&0.000000			&0.000000			&0.000000\\
&CI Q		&0.000010(3)		&0.000000			&0.000000			&0.000000			&0.000000\\
&Total		&-3714.324888(56)	&-3710.261411(31)	&-3709.423740(42)	&-3694.528520(61)	&-3696.858214(41)\\
\hline
U$^{88+}$
&PT0		&-11796.004585	    &-11785.830399	    &-11785.830399  	&-11631.048631	    &-11631.048631\\
&PT1		&-228.772670	    &-228.353746	    &-226.870923	    &-227.010654    	&-231.520289\\
&PT2 SD		&-0.814213(47)	    &-0.284411(44)	    &-0.298765(37)	    &-0.289955(50)	    &-0.239898(48)\\
&PT3 SD		&0.211751(48)		&0.002258(19)		&0.002427(22)		&0.002182(28)		&0.001467(17)\\
&CI+PT3 SD	&-0.053448(15)		&-0.000001(1)		&-0.000005(3)		&0.000008(2)		&0.000001(1)\\
&CI SD		&-0.000043(9)		&-0.000016(3)		&-0.000016(4)		&-0.000011(3)		&-0.000010(3)\\
&CI+PT3 T	&-0.001041(5)		&-0.000003(2)		&-0.000005(2)		&-0.000007(2) 		&-0.000001(1)\\
&CI T		&0.000014(2)		&-0.000002			&-0.000002(2)		&-0.000002(1)		&-0.000001(1)\\
&CI+PT3 Q	&-0.001802(7)		&0.000000			&0.000000			&0.000000			&0.000000\\
&CI Q		&0.000008(2)		&0.000000			&0.000000			&0.000000			&0.000000\\
&Total		&-12025.436029(70)	&-12014.466319(48)	&-12012.997688(43)	&-11858.347070(57)	&-11862.807362(51)\\
\hline
\end{tabular}
}
\end{table}

\begin{table}
\caption{
The DCB energies without and with the frequency-dependent Breit correction included of the ground and singly-excited $n=2$ states in Ne$^{6+}$ and U$^{88+}$, in a.u.. 
The calculations are performed for the LDF and CH screening potentials.
}
\label{table:table3}
\centering
\resizebox{\columnwidth}{!}
{
\begin{tabular}
{@{}
c c c S[table-format=-5.6(3)] 
      S[table-format=-5.6(2)] 
      S[table-format=-5.6(2)] 
      S[table-format=-5.6(3)] 
      S[table-format=-5.6(2)] 
    @{\quad}c@{}
}
\hline
\hline
\multicolumn{1}{c}{~~$Z$~~}
&\multicolumn{1}{c}{Contribution}
&\multicolumn{1}{c}{Scr. pot.}
&\multicolumn{1}{c}{$1s^22s^2\,^{1}S_{0}$}
&\multicolumn{1}{c}{$1s^22s2p\,^{3}P_{0}$}
&\multicolumn{1}{c}{$1s^22s2p\,^{3}P_{1}$}
&\multicolumn{1}{c}{$1s^22s2p\,^{1}P_{1}$}
&\multicolumn{1}{c}{$1s^22s2p\,^{3}P_{2}$}
&\multicolumn{1}{c}{$\pm(\alpha Z)^3$}\\
\hline
\multirow{4}{*}{10} &\multirow{2}{*}{DCB}        &LDF    &-110.422691(102) &-109.915159(17)&-109.913099(16)&-109.442527(115) &-109.908582(18)&\multirow{4}{*}{$\pm 0.00039$}\\
                    &                                   &CH     &-110.422734(173) &-109.915150(56)&-109.913083(37)&-109.442502(166) &-109.908561(56)    &\\
                    &\multirow{2}{*}{DCB$+$Breit$(\omega)$}   &LDF    &-110.422690(102)   &-109.915159(17)    &-109.913100(16)    &-109.442528(115)   &-109.908584(18) &\\
                    &                                   &CH     &-110.422733(173)   &-109.915150(56)    &-109.913083(37)    &-109.442503(166)   &-109.908563(56) &\\
\hline
\multirow{4}{*}{92} &\multirow{2}{*}{DCB}        &LDF    &-12025.436029(70)  &-12014.466319(48)  &-12012.997688(43)  &-11858.347070(57)  &-11862.807362(51)&\multirow{4}{*}{$\pm 0.30259$}\\
                    &                                   &CH     &-12025.439460(168) &-12014.468848(97)  &-12013.000236(92)  &-11858.350558(119) &-11862.810878(85)  &\\
                    &\multirow{2}{*}{DCB$+$Breit$(\omega)$}   &LDF    &-12025.389517(70)   &-12014.433221(48)    &-12012.964639(43)    &-11858.550937(57)   &-11863.010939(51) &\\
                    &                                   &CH     &-12025.392794(168)   &-12014.435521(97)    &-12012.966955(92)    &-11858.554549(119)   &-11863.014749(85) &\\
\hline
\end{tabular}
}
\end{table}
\pagebreak

\begin{table}
\caption{\label{table:table4}
The DCB energies for the ground and singly-excited $n=2$ states of Be-like ions with $Z=10$--$92$ supplemented with the frequency-dependent Breit correction, in a.u..
The total energies are presented for the ground state, whereas for the other states the energies relative to the ground state are given. 
The calculations are performed for the LDF potential. 
The energies are compared with the results of the previous relativistic calculations.}
\resizebox{\columnwidth}{!}
{
\begin{tabular}
{
l 
S[table-align-text-post = false, table-number-alignment=left, table-format=-6.6(2),group-separator=]
S[table-align-text-post = false, table-number-alignment=left, table-format=-3.6(2),group-separator=]
S[table-align-text-post = false, table-number-alignment=left, table-format=-3.6(2),group-separator=]
S[table-align-text-post = false, table-number-alignment=left, table-format=-4.6(2),group-separator=]
S[table-align-text-post = false, table-number-alignment=left, table-format=-4.6(2),group-separator=]
r
}
\hline
\hline
\multicolumn{1}{c}{Ion}
&\multicolumn{1}{c}{$1s^22s^2\,^1S_0$}
&\multicolumn{1}{c}{$1s^22s2p\,^3P_0$}	
&\multicolumn{1}{c}{$1s^22s2p\,^3P_1$}
&\multicolumn{1}{c}{$1s^22s2p\,^1P_1$}	
&\multicolumn{1}{c}{$1s^22s2p\,^3P_2$}
& Ref.\\
\hline
\multirow{1}{*}{Ne$^{6+}$}	&-110.42269(10)	&0.50753(10)	&0.50959(10)	&0.98016(15)	&0.51411(10)\\
	                        &			    &0.5069	        &0.5090	        &0.9794	        &0.5135&\cite{Cheng2008PRABe-like}\\
	                        &		        &0.5063	        &0.5084	        &0.9732	        &0.5129&\cite{Safronova1996PRABe-like}\\
\multirow{1}{*}{Ar$^{14+}$} &-379.38562(6)  &1.04647(7)     &1.07912(7)     &2.06504(9)     &1.15558(7)\\
	                        &-379.3853		&1.0465			&1.0791		    &2.0653			&1.1556&\cite{Yerokhin2015PhysScr_Be-likeAr}\\
	                        &				&1.0462		    &1.0788			&2.0615			&1.1553& \cite{Safronova1996PRABe-like}\\
\multirow{1}{*}{Fe$^{22+}$} &-812.59026(6)  &1.60359(6)     &1.74386(6)     &3.44430(9)     &2.16495(6)&\\
                            &-812.5898		&1.6037			&1.7440			&3.4448			&2.1653&\cite{Yerokhin2014PRA_Be-likeFe}\\
                            &				&1.6036			&1.7439			&3.4448			&2.1649&\cite{Cheng2008PRABe-like}\\
                            &				&1.6034			&1.7436			&3.4419			&2.1647&\cite{Safronova1996PRABe-like}\\
\multirow{1}{*}{Cu$^{25+}$}	&-1018.31580(6)	&1.82046(6)     &2.02396(6)     &4.12200(9)     &2.72516(6)\\
                            &				&1.8203			&2.0236			&4.1199			&2.7248			&\cite{Safronova1996PRABe-like}\\
\multirow{1}{*}{Zn$^{26+}$} &-1092.22806(9) &1.89411(11)    &2.12087(12)	&4.37677(14)    &2.94268(11)\\
                            &			    &1.8941			&2.1209         &4.3772			&2.9426			& \cite{Cheng2008PRABe-like}\\
                            &				&1.89389		&2.12053		&4.37467		&2.94233	&\cite{Safronova1996PRABe-like}\\
\multirow{1}{*}{Kr$^{32+}$} &-1592.72137(6)	&2.35438(7)     &2.73476(7)     &6.31245(9)     &4.66589(7)\\
                            &	            &2.3544	        &2.7349	        &6.3125	        &4.6658       &\cite{Cheng2008PRABe-like}\\
                            &	            &2.3566	        &2.7365	        &6.3134	        &4.6656	      &\cite{Safronova2000MolPhys_Be-like}\\	
                            &	            &2.3542	        &2.7343	        &6.3105	        &4.6653	        &\cite{Safronova1996PRABe-like}\\
\multirow{1}{*}{Mo$^{38+}$} &-2193.53880(7) &2.85758(8)     &3.39847(8)     &9.21628(10)    &7.35801(8)\\	
	                        &	            &2.8577	        &3.3988	        &9.2163	        &7.3579	        & \cite{Cheng2008PRABe-like}\\
	                        &	            &2.8574	        &3.3980	        &9.2144	        &7.3572	        &\cite{Safronova1996PRABe-like}\\
\multirow{1}{*}{Xe$^{50+}$}&-3714.32145(7)  &4.06066(7)     &4.89818(7)     &19.77947(8)    &17.44970(7)\\
	                       &-3714.320&	    &	            &	            &	            &\cite{Malyshev2014PRA_Be-like_GS}\\
	                       &	            &4.0601	        &4.8976	        &19.7766	    &17.4474        &\cite{Safronova1996PRABe-like}\\
\multirow{1}{*}{Nd$^{56+}$}&-4646.20449(5)  &4.79683(6)     &5.76626(6)     &28.64454(7)    &26.04788(6)    &\\
	                       &	            &4.7971	        &5.7660	        &28.6411	    &26.0449        &\cite{Safronova1996PRABe-like}\\
\hline
\end{tabular}
}
\end{table}

\begin{table}
Table~\ref{table:table4}. (\textit{Continued.})
\resizebox{\columnwidth}{!}
{
\begin{tabular}
{
l 
S[table-align-text-post = false, table-number-alignment=left, table-format=-6.6(2),group-separator=]
S[table-align-text-post = false, table-number-alignment=left, table-format=-3.6(2),group-separator=]
S[table-align-text-post = false, table-number-alignment=left, table-format=-3.6(2),group-separator=]
S[table-align-text-post = false, table-number-alignment=left, table-format=-4.6(2),group-separator=]
S[table-align-text-post = false, table-number-alignment=left, table-format=-4.6(2),group-separator=]
r
}
\hline
\hline
Ion
&\multicolumn{1}{c}{$^1S_0$}
&\multicolumn{1}{c}{$^3P_0$}
&\multicolumn{1}{c}{$^3P_1$}
&\multicolumn{1}{c}{$^1P_1$}
&\multicolumn{1}{c}{$^3P_2$}
& Ref.\\
\hline
\multirow{1}{*}{Yb$^{66+}$}&-6481.15683(5)  &6.28448(6)     &7.45012(6)     &51.59172(8)    &48.49943(6)    \\
	                       &	            &6.2747	        &7.4392	        &51.5812	    &48.4974        &\cite{Safronova1996PRABe-like}\\
\multirow{1}{*}{Hg$^{76+}$}&-8714.07758(6)  &8.18212(7)     &9.51208(7)     &89.59987(9)    &85.93975(7)    &\\
	                       &	            &8.1838	        &9.5125	        &89.5902	    &85.9295	    &\cite{Safronova1996PRABe-like}\\
\multirow{1}{*}{Bi$^{79+}$}&-9471.84553(6)  &8.83780(7) &10.20951(7)    &105.06042(9)   &101.21364(7)\\
	                       &	            &8.8398     &10.2101	    &105.049        &101.202    &\cite{Safronova1996PRABe-like}\\
\multirow{1}{*}{Th$^{86+}$}&-11418.88796(6) &10.47339(7)&11.92476(7)    &150.80430(8)   &146.48804(7)	 \\
	                       &	            &10.4728	&11.9244	    &150.801	    &146.483	&\cite{Cheng2008PRABe-like}\\
	                       &	            &10.4748	&11.9241	    &150.786	    &146.469	&\cite{Safronova1996PRABe-like}\\
\multirow{1}{*}{U$^{88+}$}  &-12025.38952(7)&10.95630(8)&12.42488(8)    &166.83858(9)   &162.37858(9)\\
                            &-12025.387	    &           &	            &	            &           &\cite{Malyshev2014PRA_Be-like_GS}\\
                            &	            &10.9594	&12.4286	    &166.840	    &162.379    &\cite{Cheng2008PRABe-like}\\          
                            &	            &10.9617	&12.4280	    &166.823	    &162.362    &\cite{Safronova1996PRABe-like}\\ 
\hline
\hline	
\end{tabular}
}
\end{table}
\pagebreak

\begin{table}
\caption{\label{table:table5} 
The total binding energies for berylliumlike ions with $Z=10-92$, in a.u.. 
The DCB energy, the frequency-dependent Breit correction, QEDMOD and nuclear recoil corrections are shown separately. 
For the ground state the total energies are presented, whereas for all other states the energies relative to the ground state are given. 
The comparison with the recent relativistic calculations \cite{Cheng2008PRABe-like, Malyshev2014PRA_Be-like_GS, Yerokhin2015PhysScr_Be-likeAr, Yerokhin2014PRA_Be-likeFe}, the data from the NIST Atomic Spectra Database~\cite{NIST_ASD}, and the available experimental values is presented.
}
%
\resizebox{\columnwidth}{!}{
\sisetup{
table-space-text-post = \textsuperscript{\emph{a}}
}
\begin{tabular}
{
c
c
S[table-format=-6.6(6),parse-numbers=false, group-separator=]
S[table-format=-9.6,group-separator=]
S[table-format=-2.6(2),group-separator=]
S[table-format=-2.6,group-separator=]
S[table-align-text-post = false, table-format=-6.6(3),group-separator=]
S[table-align-text-post = false, table-number-alignment=left, table-format=-2.5(2),group-separator=]
S[table-align-text-post = false, table-number-alignment=left, table-format=-6.5(2),group-separator=]
S[table-align-text-post = false, table-number-alignment=left, table-format=3.4(2),group-separator=]
                 }
\hline\hline
\multicolumn{1}{c}{  ~~$Z$~~ } 
& \multicolumn{1}{c}{  Term } 
& \multicolumn{1}{c}{ ~~~~~~~~~~~~~DCB }  
& \multicolumn{1}{c}{ ~~~~~~~Breit$(\omega)$ } 
& \multicolumn{1}{c}{ QEDMOD } 
& \multicolumn{1}{c}{ Recoil } 
& \multicolumn{1}{c}{ Total }
& \multicolumn{1}{c}{ Other theory} 
& \multicolumn{1}{c}{ NIST }
& \multicolumn{1}{c}{ Experiment }
\\
\hline
10	&$^1S_0$		&-110.42269(10)(0)	&0.00000	&0.01096(21) 	&0.00305    &-110.40868(45)	&														&-110.4087(5)&		\\
	&$^3P_0$		&0.50753(10)(0)		&0.00000	&-0.00044(26)	&-0.00013	&0.50696(62)		&0.5069 \textsuperscript{\textit{a}}		&0.50691(2)	&		\\
	&$^3P_1$		&0.50959(10)(0)		&0.00000	&-0.00043(26)	&-0.00013	&0.50903(62)		&0.5090 \textsuperscript{\textit{a}}		&0.50898(2)	&		\\
	&$^3P_2$		&0.51411(10)(0)		&0.00000	&-0.00046(26)	&-0.00013	&0.51356(62)		&0.5135 \textsuperscript{\textit{a}}		&0.51351(2)	&		\\
	&$^1P_1$		&0.98016(15)(0)		&0.00000	&-0.00041(27)	&-0.00014	&0.97956(63)		&0.9794 \textsuperscript{\textit{a}}		&0.979392(6)&	\\
18	&$^1S_0$		&-379.38563(6)(1)	&0.00001	&0.0880(13)		&0.00577	&-379.2919(25)  &-379.2918(8)\textsuperscript{\textit{b}}&	&\\
 	&				&					&			&				&			&				&-379.2909(15)\textsuperscript{\textit{c}}&	&\\
 	&$^3P_0$		&1.04649(7)(0)		&-0.00001   &-0.0042(16)	&-0.00028	&1.0420(35)		&1.0422(4)\textsuperscript{\textit{b}}  &   &1.0419(9)\textsuperscript{\textit{d}}\\
 	&				&					&			&				&			&				&1.0419\textsuperscript{\textit{a}}		&	&\\		
 	&$^3P_1$		&1.07914(7)(0)		&-0.00002   &-0.0041(16)	&-0.00028	&1.0747(35)		&1.0749(6)\textsuperscript{\textit{b}}	&	&1.07466(5)\textsuperscript{\textit{d}}\\
 	&				&					&			&				&			&				&1.0747\textsuperscript{\textit{a}}		&	&	\\
 	&$^3P_2$		&1.15563(7)(0)		&-0.00005   &-0.0039(17)	&-0.00028	&1.1514(35)		&1.1516(10)\textsuperscript{\textit{b}} &	&1.1513(6)\textsuperscript{\textit{d}}\\
 	&				&					&			&				&			&				&1.1513\textsuperscript{\textit{a}}	    &	&	\\		
 	&$^1P_1$		&2.06508(9)(0)		&-0.00004	&-0.0042(17)	&-0.00030	&2.0605(35)	    &2.0611(4)\textsuperscript{\textit{b}}	&	&2.06048(9)\textsuperscript{\textit{d}}\\
 	&	            &	                &	        &	            &	        &	            &2.0604\textsuperscript{\textit{a}}	    &	&\\				
26	&$^1S_0$	&-812.59034(6)(4)	&0.00008	&0.3109(29)	&0.00790	&-812.2714(74)	&-812.2695(30)\textsuperscript{\textit{e}}		&-812.24(7)	&\\
	&			&					&			&			&			&				&-812.2688(17)\textsuperscript{\textit{c}}		&			&\\
    &$^3P_0$	&1.60367(6)(0)	    &-0.00008	&-0.0162(36)&-0.00041	&1.587(10)	    &1.5875(45)\textsuperscript{\textit{e}}		    &1.5865(15)	&\\
	&           &	                &	        &	        &           &               &1.5870\textsuperscript{\textit{a}}	            &	        &\\
	&$^3P_1$	&1.74399(6)(0)	    &-0.00014	&-0.0159(36)&-0.00041	&1.728(10)	    &1.7280(45)\textsuperscript{\textit{e}}		    &1.7274(3)	&\\
	&	        &	                &           &	        &	        &	            &1.7275\textsuperscript{\textit{a}}	            &	        &\\
	&$^3P_2$	&2.16534(6)(0)	    &-0.00038	&-0.0148(37)&-0.00041	&2.150(10)	    &2.1505(45)\textsuperscript{\textit{e}}		    &2.1496(3)	&\\
	&	        &	                &	        &           &           &	            &2.1496\textsuperscript{\textit{a}}	            &	        &\\
    &$^1P_1$	&3.44463(9)(0)	    &-0.00033	&-0.0156(38)&-0.00042	&3.428(10)	    &3.4290(45)\textsuperscript{\textit{e}}		    &3.4282(3)	&\\
	&	        &	                &	        &	        &	        &               &3.4282\textsuperscript{\textit{a}}	            &	        &\\
 29	&$^1S_0$	&-1018.31594(6)(1)	&0.00014	&0.4501(40)		&0.00878	&-1017.857(10)	&					                    &-1017.81(8)&\\
 	&$^3P_0$	&1.82060(6)(0)		&-0.00014	&-0.0240(50)	&-0.00046	&1.796(14)		&1.7947\textsuperscript{\textit{a}}		&1.795(5)	&\\
 	&$^3P_1$	&2.02418(6)(0)		&-0.00022	&-0.0237(50)	&-0.00046	&2.000(14)		&2.0001\textsuperscript{\textit{a}}		&2.0001(5)	&\\
 	&$^3P_2$	&2.72584(6)(0)		&-0.00068	&-0.0218(50)	&-0.00047	&2.703(14)		&2.7032\textsuperscript{\textit{a}}		&2.7033(5)	&\\
 	&$^1P_1$	&4.12261(9)(0)		&-0.00060	&-0.0228(51)	&-0.00048	&4.099(14)		&4.0979\textsuperscript{\textit{a}}		&4.0979(5)	&\\
30	&$^1S_0$	&-1092.22823(6)(1)	&0.00017	&0.5047(44)		&0.00926	&-1091.714(11) &-1091.7106(17)\textsuperscript{\textit{c}}	    &-1091.67(9)&\\
 	&$^3P_0$	&1.89427(6)(0)		&-0.00016	&-0.0271(54)	&-0.00048 	&1.866(16)		&1.8663\textsuperscript{\textit{a}}				&           &\\
 	&$^3P_1$	&2.12112(6)(0)		&-0.00026	&-0.0268(54)	&-0.00049	&2.094(16)		&2.0934\textsuperscript{\textit{a}}				&	        &\\
 	&$^3P_2$	&2.94350(6)(0)		&-0.00081	&-0.0246(55)	&-0.00050	&2.918(16)		&2.9172\textsuperscript{\textit{a}}				&	        &\\
 	&$^1P_1$	&4.37750(9)(0)		&-0.00073	&-0.0257(56)	&-0.00051	&4.351(16)		&4.3515\textsuperscript{\textit{a}}				&	        &\\
\hline
\end{tabular}
\footnotetext{$^{a}$ Cheng \textit{et al.} \cite{Cheng2008PRABe-like}.}
\footnotetext{$^{b}$ Yerokhin \textit{et al.} \cite{Yerokhin2015PhysScr_Be-likeAr}.}
\footnotetext{$^{c}$ Malyshev \textit{et al.} \cite{Malyshev2014PRA_Be-like_GS}.}
\footnotetext{$^{d}$ Edl{\'e}n  \cite{Edlen1983PhysScr_Be-like}.}
\footnotetext{$^{e}$ Yerokhin \textit{et al.} \cite{Yerokhin2014PRA_Be-likeFe}.}
}
\end{table}

\newpage

\begin{table}
Table~\ref{table:table5}. (\textit{Continued.})
%
\resizebox{\columnwidth}{!}{
\begin{tabular}
{
        c
        c
        S[table-format=-6.6(6),parse-numbers=false, group-separator=]
        S[table-format=-9.6,group-separator=]
        S[table-format=-2.6(2),group-separator=]
        S[table-format=-2.6,group-separator=]
        S[table-align-text-post = false, table-format=-6.6(3),group-separator=]
        S[table-align-text-post = false, table-number-alignment=left, table-format=-2.5(2),group-separator=]
        S[table-align-text-post = false, table-number-alignment=left, table-format=-6.5(2),group-separator=]
        S[table-align-text-post = false, table-number-alignment=left, table-format=3.4(2),group-separator=]
                 }
\hline\hline
\multicolumn{1}{c}{  ~~$Z$~~ } 
& \multicolumn{1}{c}{  Term } 
& \multicolumn{1}{c}{ ~~~~~~~~~~~~~DCB }  
& \multicolumn{1}{c}{ ~~~~~~~Breit$(\omega)$ } 
& \multicolumn{1}{c}{ QEDMOD } 
& \multicolumn{1}{c}{ Recoil } 
& \multicolumn{1}{c}{ Total } 
& \multicolumn{1}{c}{ Other theory} 
& \multicolumn{1}{c}{ NIST } 
& \multicolumn{1}{c}{ Experiment } 
\\
\hline
36	&$^1S_0$	&-1592.72180(6)(3)	&0.00044	&0.9310(73)		&0.01024	&-1591.780(20)	    &-1591.7746(18)\textsuperscript{\textit{c}} &-1591.7(2) &\\
	&$^3P_0$	&2.35479(7)(0)		&-0.00040	&-0.0520(92)	&-0.00054	&2.302(27)			&2.3014\textsuperscript{\textit{a}}			&2.303(4)   &\\
	&$^3P_1$	&2.73530(7)(0)		&-0.00054	&-0.0517(92)	&-0.00054	&2.682(27)			&2.6822\textsuperscript{\textit{a}}			&2.6827(4)  &\\
	&$^3P_2$	&4.66800(7)(0)		&-0.00211	&-0.0467(92)	&-0.00056	&4.619(27)			&4.6174\textsuperscript{\textit{a}}			&4.6175(4)  &\\
	&$^1P_1$	&6.31444(9)(0)		&-0.00198	&-0.0482(94)	&-0.00057	&6.264(27)			&6.2641\textsuperscript{\textit{a}}			&6.2625(4)  &\\
42	&$^1S_0$	&-2193.53976(7)(7)	&0.00096	&1.558(12)	    &0.01201	&-2191.969(31)	    &-2191.9603(20)\textsuperscript{\textit{c}}	&-2191.9(2)&\\
	&$^3P_0$	&2.85844(8)(0)		&-0.00086	&-0.090(14)	    &-0.00063	&2.767(43)	        &2.7662\textsuperscript{\textit{a}}	        &	        &\\
 	&$^3P_1$	&3.39948(8)(0)		&-0.00102	&-0.090(14)	    &-0.00064	&3.308(43)	        &3.3076\textsuperscript{\textit{a}}	        &	        &\\
 	&$^3P_2$	&7.36268(8)(0)		&-0.00467	&-0.080(14)	    &-0.00067	&7.277(43)	        &7.2757\textsuperscript{\textit{a}}	        &	        &\\
 	&$^1P_1$	&9.22082(10)(0)		&-0.00455	&-0.082(15)	    &-0.00068	&9.134(43)	        &9.1337\textsuperscript{\textit{a}}	        &	        &\\
54	&$^1S_0$	&-3714.32489(6)(60) &0.00343	&3.600(25)	    &0.01483	&-3710.707(66)	&-3710.6864(31)\textsuperscript{\textit{c}}	&-3710.3(2)&\\
 	&$^3P_0$	&4.06348(7)(4)      &-0.00282	&-0.218(31)	    &-0.00076	&3.842(92)	    &3.8393\textsuperscript{\textit{a}}	&   &\\
 	&$^3P_1$	&4.90115(7)(4)      &-0.00296	&-0.218(31)	    &-0.00077	&4.679(92)	    &4.6775\textsuperscript{\textit{a}}	&	&4.6770(17)\textsuperscript{\textit{f}}\\
 	&$^3P_2$	&17.46667(7)(4)	    &-0.01697	&-0.192(31)	    &-0.00085	&17.256(92)	    &17.252\textsuperscript{\textit{a}}	&	&17.2529(30)\textsuperscript{\textit{f}}\\
 	&$^1P_1$	&19.79637(8)(4)     &-0.01690	&-0.195(31)	    &-0.00086	&19.584(92)	    &19.583\textsuperscript{\textit{a}}	&	&19.5801(6)\textsuperscript{\textit{f}}\\
60	&$^1S_0$	&-4646.21031(5)(92) &0.00582	&5.125(34)		&0.01704	&-4641.062(91)	&-4641.0341(44)\textsuperscript{\textit{c}}	&-4640.7(2)	&\\
 	&$^3P_0$	&4.80132(6)(4)      &-0.00449	&-0.315(43) 	&-0.00085	&4.48(13)	    &4.4766\textsuperscript{\textit{a}}	&	&	\\
	&$^3P_1$	&5.77088(6)(4)		&-0.00462	&-0.316(43) 	&-0.00085	&5.45(13)	    &5.4466\textsuperscript{\textit{a}}	&	&	\\
 	&$^3P_2$	&26.07691(6)(4)		&-0.02902	&-0.279(43) 	&-0.00099	&25.77(13)	    &25.7619\textsuperscript{\textit{a}}&	&	\\
 	&$^1P_1$	&28.67355(7)(4)		&-0.02901	&-0.282(44) 	&-0.00099	&28.36(13)	    &28.3600\textsuperscript{\textit{a}}&	&	\\
70	&$^1S_0$	&-6481.16933(5)(420)&0.01250	&8.64(10)	    &0.01886	&-6472.50(17)   &-6472.4491(92)\textsuperscript{\textit{c}}	&-6472(2)&\\
	&$^3P_0$	&6.29282(6)(28)		&-0.00834	&-0.54(13)	    &-0.00086	&5.74(23)		&5.7331\textsuperscript{\textit{a}}	&	    &\\
 	&$^3P_1$	&7.45858(6)(28)		&-0.00845	&-0.55(13)	    &-0.00086	&6.90(23)		&6.8994\textsuperscript{\textit{a}}	&	    &\\
 	&$^3P_2$	&48.56289(6)(28)	&-0.06345	&-0.48(13)	    &-0.00111	&48.01(23)		&48.0049\textsuperscript{\textit{a}}&	    &\\
 	&$^1P_1$	&51.65526(8)(28)	&-0.06354	&-0.49(13)	    &-0.00110	&51.10(23)		&51.0988\textsuperscript{\textit{a}}&	    &\\
80	&$^1S_0$	&-8714.10155(6)(993)&0.02397	&13.72(9)	    &0.02089	&-8700.33(22)	&-8700.258(19)\textsuperscript{\textit{c}}	&-8700(4)	&\\
	&$^3P_0$	&8.19475(7)(68)		&-0.01263	&-0.88(12)	    &-0.00080	&7.30(30)		&7.2854\textsuperscript{\textit{a}}	&	&\\
 	&$^3P_1$	&9.52480(7)(68)		&-0.01273	&-0.88(12)	    &-0.00080	&8.63(30)		&8.6161\textsuperscript{\textit{a}}	&	&\\
 	&$^3P_2$	&86.06413(7)(68)	&-0.12438	&-0.79(12)	    &-0.00124	&85.15(30)		&85.1225\textsuperscript{\textit{a}}&	&\\
 	&$^1P_1$	&89.72446(9)(68)	&-0.12459	&-0.80(12)	    &-0.00122	&88.80(30)	    &88.7853\textsuperscript{\textit{a}}&	&\\ 
83	&$^1S_0$	&-9471.87413(6)(1331)	&0.02860	&15.63(10)	&0.02154	&-9456.20(28)   &			                        &-9456(5)   &\\
	&$^3P_0$	&8.85142(7)(93)			&-0.01362	&-1.01(13)	&-0.00075	&7.83(39)		&7.8129\textsuperscript{\textit{a}}	&	&	\\
 	&$^3P_1$	&10.22321(7)(93)		&-0.01370	&-1.01(13)	&-0.00076	&9.20(39)		&9.1852\textsuperscript{\textit{a}}	&	&	\\
 	&$^3P_2$	&101.36324(7)(93)		&-0.14961	&-0.91(13)	&-0.00128	&100.30(39)	    &100.273\textsuperscript{\textit{a}}&	&	\\
 	&$^1P_1$	&105.21028(9)(93)		&-0.14986 	&-0.92(13)	&-0.00125	&104.14(39)	    &104.123\textsuperscript{\textit{a}}&	&	\\ 			
90	&$^1S_0$	&-11418.92993(6)(5194)	&0.04198	&20.90(14)	&0.02197	&-11397.97(32) &-11397.876(63)\textsuperscript{\textit{c}}			&-11397(9)	&\\
	&$^3P_0$	&10.48746(8)(370)		&-0.01407	&-1.35(18)	&-0.00053	&9.12(44)		&9.0940\textsuperscript{\textit{a}}	&	&	\\
 	&$^3P_1$	&11.93889(7)(370)		&-0.01413	&-1.35(18)	&-0.00054	&10.57(44)	    &10.5459\textsuperscript{\textit{a}}	&	&	\\
 	&$^3P_2$	&146.71223(7)(370)		&-0.22419	&-1.25(18)	&-0.00131	&145.23(44)	    &145.195\textsuperscript{\textit{a}}	&	&	\\
 	&$^1P_1$	&151.02878(8)(370)		&-0.22449	&-1.26(18)	&-0.00125	&149.54(44)	    &149.514\textsuperscript{\textit{a}}	&	&	\\
92	&$^1S_0$    &-12025.43603(7)(1429)	&0.04651	&22.64(16)	&0.02201	&-12002.74(34)  &-12002.645(48)\textsuperscript{\textit{c}} &-12003(10)&\\
	&$^3P_0$	&10.96971(8)(102)		&-0.01341	&-1.46(20)	&-0.00043	&9.51(47)		&9.4653\textsuperscript{\textit{a}}	    &	&	\\
 	&$^3P_1$	&12.43834(8)(102)		&-0.01346	&-1.46(20)	&-0.00045	&10.97(47)	    &10.9343\textsuperscript{\textit{a}}	&	&	\\
 	&$^3P_2$	&162.62867(9)(102)	    &-0.25009	&-1.36(20)	&-0.00131	&161.02(47)	    &160.972\textsuperscript{\textit{a}}	&	&	\\
 	&$^1P_1$	&167.08896(9)(102)		&-0.25038	&-1.37(20)	&-0.00125	&165.47(47)	    &165.433\textsuperscript{\textit{a}}	&	&	\\
\hline
\end{tabular}
}
\footnotetext{$^{a}$ Cheng \textit{et al.} \cite{Cheng2008PRABe-like}.}
\footnotetext{$^{c}$ Malyshev \textit{et al.} \cite{Malyshev2014PRA_Be-like_GS}.}
\footnotetext{$^{f}$ Bernhardt \textit{et al.} \cite{Bernhardt2015JPhysB_Be-like_Xe}.}
\end{table}

\clearpage
\bibliography{mybib} 
\end{document}